\documentclass{article}
\usepackage{spconf,amsmath,graphicx}

\usepackage{graphicx}

\usepackage{blindtext,letltxmacro,xcolor,xparse}
\usepackage{subcaption}
\usepackage{booktabs}
\usepackage{algorithmic}
\usepackage{hyperref}
\usepackage{mathtools}
\usepackage{amsfonts}
\usepackage{amsmath}
\usepackage{enumitem}
\usepackage{float}
\usepackage{mathtools}
\usepackage{pgfplots}
\usepackage{cite}
\usepackage{algorithm}

\LetLtxMacro{\blindtextblindtext}{\blindtext}
\LetLtxMacro{\blindtextBlindtext}{\Blindtext}

\RenewDocumentCommand{\blindtext}{O{\value{blindtext}}}{%
  \begingroup\color{gray}\blindtextblindtext[#1]\endgroup
}
\RenewDocumentCommand{\Blindtext}{O{\value{blindtext}}O{\value{Blindtext}}}{%
  \begingroup\color{gray}\blindtextBlindtext[#1][#2]\endgroup
}

\makeatletter
\renewenvironment{description}%
               {\list{}{\leftmargin=10pt % <------- Adjust this length
                        \labelwidth\z@ \itemindent-\leftmargin
                        }}%
               {\endlist}
\makeatother

\DeclareMathOperator*{\argmax}{argmax}

\usepackage{tikz}
\usetikzlibrary{positioning}
\usetikzlibrary{fit}
\usetikzlibrary{backgrounds}
\usetikzlibrary{calc}
\usetikzlibrary{shapes}
\usetikzlibrary{arrows}
\usetikzlibrary{decorations}
\tikzset{%
  block/.style={rectangle,draw=black,fill=white,thick,minimum height=1cm,inner sep=10pt, align=center},
  arrow/.style={->,thick},
  }

\newcommand\blfootnote[1]{%
  \begingroup
  \renewcommand\thefootnote{}\footnote{#1}%
  \addtocounter{footnote}{-1}%
  \endgroup
}

\ninept
\title{Integration of variational autoencoder and spatial clustering \\
for adaptive multi-channel neural speech separation}
\name{Katerina Zmolikova$^1$, Marc Delcroix$^2$, Luk\'{a}\v{s} Burget$^1$, Tomohiro Nakatani$^2$, Jan "Honza" \v{C}ernock\'{y}$^1$}
\address{$^1$Brno University of Technology, Faculty of IT, IT4I Centre of Excellence \\ $^2$NTT Communication Science Laboratories, NTT Corporation, Kyoto, Japan}

\begin{document}
%\ninept
%
\maketitle
\begin{abstract}
In this paper, we propose a method combining variational autoencoder model of speech with a spatial clustering approach for multi-channel speech separation. The advantage of integrating spatial clustering with a spectral model was shown in several works. As the spectral model, previous works used either factorial generative models of the mixed speech or discriminative neural networks. In our work, we combine the strengths of both approaches, by building a factorial model based on a generative neural network, a variational autoencoder. By doing so, we can exploit the modeling power of neural networks, but at the same time, keep a structured model. Such a model can be advantageous when adapting to new noise conditions as only the noise part of the model needs to be modified. We show experimentally, that our model significantly outperforms previous factorial model based on Gaussian mixture model (DOLPHIN), performs comparably to integration of permutation invariant training with spatial clustering, and enables us to easily adapt to new noise conditions. 
\blfootnote{The work was partly supported by Czech National Science Foundation (GACR) project NEUREM3 No. 19-26934X, and by Czech Ministry of Education, Youth and Sports from project no. LTAIN19087 "Multi-linguality in speech technologies".}
\end{abstract}
\begin{keywords}
Multi-channel speech separation, variational autoencoder, spatial clustering, DOLPHIN
\end{keywords}
\vspace{-0.3em}
\section{Introduction}
\label{sec:intro}
\vspace{-0.5em}
Multi-channel speech separation is a problem of great importance for applications of speech technologies, such as personal devices or transcription of meetings. A popular class of approaches to tackle this problem is based on spatial clustering \cite{ito:cacgmm,sawada:undetermined,vu2010blind,mandel09a}. In these approaches, a mixture model of spatial features (e.g. phase differences \cite{mandel09a} or normalized observation vectors \cite{ito:cacgmm}) is built. By inferring the affiliations of observed spatial features to the mixture components, we obtain time-frequency masks, which can then be used for separating the signal. The spatial clustering approaches have proven their strength in many tasks, e.g. recent CHiME challenges \cite{boeddeker2018front,watanabe2020chime}.

The spatial clustering approaches make use of the spatial properties of the speech signal, however, they do not use any information about the spectral structure of speech. To overcome this drawback, methods integrating the spatial model with a spectral model have been proposed \cite{nakatani2013dominance,reyes:fhmm,nakatani2017integrating,drude:integration}. We can observe two main directions of spectral models for such integration: first, factorial generative models of the mixed speech \cite{nakatani2013dominance,reyes:fhmm} and second, discriminative neural networks \cite{nakatani2017integrating,drude:integration}.

In \textit{factorial generative models}, models of clean speech and noise signals are trained and combined into a factorial model of the mixture. In the past, such models have been applied also to single-channel separation \cite{rennie2010single} and have been integrated with spatial models for multi-channel case \cite{nakatani2013dominance,reyes:fhmm}. Notably, in DOminance based Locational and Power-spectral cHaracteristics INtegration (DOLPHIN) \cite{nakatani2013dominance}, Gaussian mixture models are used to model speech and noise signals, and the index of the dominant source is shared with a spatial model, to provide joint inference in both models. Factorial models have demonstrated great potential for separation at early challenges \cite{delcroix2011speech,hershey2010super}. A clear advantage of the factorial model is the explicit structure, which enables us to modify different parts of the model. For instance, if noise conditions in the recordings change, we can only modify the noise model, without changing the other parts, which enables to adapt these models easily to the recording conditions. On the other hand, models such as Gaussian mixture models are not flexible enough to capture the structure of speech well, as has become apparent in many speech tasks.

More recently, approaches based on \textit{discriminative neural networks} have been applied. The success of approaches such as deep clustering \cite{hershey2016deep} or permutation invariant training (PIT) \cite{yu2017permutation,luo2019conv} started a new era in speech separation, where neural networks are considered state-of-the-art. In these approaches, a neural network is trained to map from the mixed signal to the individual sources (represented either directly, as time-frequency masks, or embeddings). The neural network-based approaches have been also combined with spatial clustering in \cite{nakatani2017integrating,drude:integration}. There, the outputs of the network are used as priors on the mixture weights in a spatial model. While the neural networks offer great modeling power, they suffer from problems in mismatched conditions. In contrast with the factorial models, we have no insight into how the neural network works with the speech and noise components of the signal. To adapt to different noise conditions, we thus need to retrain the full model on in-domain data.

To combine the strengths of both stated directions, we here propose to use a neural network as a generative model of the speech signal in a factorial model. For this, we use the variational autoencoder (VAE) \cite{kingma2013auto,rezende2014stochastic}, which has been previously successfully used to model speech \cite{hsu2017unsupervised,gburrek2019unsupervised}. We train VAE on clean single-speaker speech and create a simple Gaussian model of noise. Using the lifted-max-model of interaction \cite{rennie2010single}, we combine the model of individual sources into a model of the mixture. As in DOLPHIN, we integrate the factorial model with a spatial model through the dominance index. At inference, to estimate the time-frequency masks of each source, we iteratively infer the time-frequency masks and the spectral properties of the signals, represented as latent variables.

Experimentally, we show that our proposed model 1) is stronger than DOLPHIN with GMM as a spectral model, 2) provides comparable results to PIT combined with spatial clustering on matched conditions, and 3) can adapt to different noise conditions by modifying the noise model. The last property can be useful for example in the scenario of long recordings, where we could progressively adapt the noise model with more incoming data.

\section{Problem formulation}
In this paper, we process an observed multi-channel mixture of speech from multiple speakers and noise with the goal of separating the speech signals. 
In the time-domain, the mixture signal at the $m$-th microphone can be expressed as,
\begin{equation}
    y_m[\tau] = \sum_{i} x_{m,i}[\tau] + n_m[\tau],
\end{equation}
where $\tau$ is the time index, and $y_m[\tau]$ $x_{m,i}[\tau]$ $n_m[\tau]$ are the time domain signals of the mixture, the speech sources and the noise respectively. In the following we consider the case of two speakers, i.e. $i\in\{1,2\}$ and work with all signals in Short-time Fourier transform (STFT) domain. Let us introduce following notations which will be used throughout the paper:
\begin{itemize}[leftmargin=1.5em]
\itemsep-0.2em
\item $\mathcal{Y}_{m,t,f}$: STFT coefficient of the observed mixed signal at $m$-th microphone for time frame $t$ and frequency bin $f$.
\item $\boldsymbol{\mathcal{Y}}_{t,f} = [\mathcal{Y}_{m,t,f}]_{m=1..M}^\top $: vector of STFT coefficients of the observed mixed signal at all microphones, where $M$ is the number of microphones.
\item $\mathbf{\Psi} = [\boldsymbol{\psi}_{t}]_{t=1..T}$, with $\boldsymbol{\psi}_t = [\boldsymbol{\psi}_{t,f}]^\top_{f=1..F}$: tensor of spatial features at all channels and all time-frequency bins $(t,f)$. The specific form of spatial features we use, is introduced in Eq. (\ref{eq:spat}).
\item $y_{t,f}$, $x^{(i)}_{t,f}$ and $n_{t,f}$: the log-magnitude of STFT coefficient at the first microphone and the $(t,f)$ time-frequency bin of the observed mixture signal, the single-speaker speech signal of speaker $i\in\{1,2\}$ and the noise signal, respectively.
%\item $y_{t,f}  = \log |\mathcal{Y}_{m=0,t,f} |$: log-magnitude of STFT coefficients of observed mixed signal at the first microphone at the $(t,f)$ time-frequency bin.
\item $\mathbf{Y} = [\mathbf{y}_{t}]_{t=1..T}$, with $\mathbf{y}_t = [y_{t,f}]^\top_{f=1..F}$: matrix of log-magnitude of STFT coefficients of the observed mixed signal at the first microphone and at all time-frequency bins $(t,f)$.
%\item $x^{(i)}_{t,f}$: log-magnitude of STFT the coefficient of single-speaker speech signal of speaker $i\in\{1,2\}$ at first microphone for the $(t,f)$ time-frequency bin.
\item $\mathbf{X}^{(i)} = [\mathbf{x}^{(i)}_{t}]_{t=1..T}$, with $\mathbf{x}^{(i)}_t = [x^{(i)}_{t,f}]^\top_{f=1..F}$: matrix of log-magnitude of STFT coefficients of single-speaker speech signal of speaker $i\in\{1,2\}$ at the first microphone and  at all time-frequency bins $(t,f)$. We omit the $i$ index in cases where we refer to single-speaker speech in general, not in the mixture.
\item $\mathbf{N} = [\mathbf{n}_{t}]_{t=1..T}$, with $\mathbf{n}_t = [n_{t,f}]^\top_{f=1..F}$: matrix of log-magnitude of STFT coefficients of noise signal at the first microphone and  at all time-frequency bins $(t,f)$.
\end{itemize}

The aim of all conventional and proposed approaches in this paper is to estimate time-frequency (T-F) masks for all sources. These masks are then used to extract all speech sources using mask-based beamforming \cite{heymann2016neural}.

\section{Conventional approaches}
\label{sec:prev_models}

\begin{figure}
    \centering
        \includegraphics[width=0.8\linewidth]{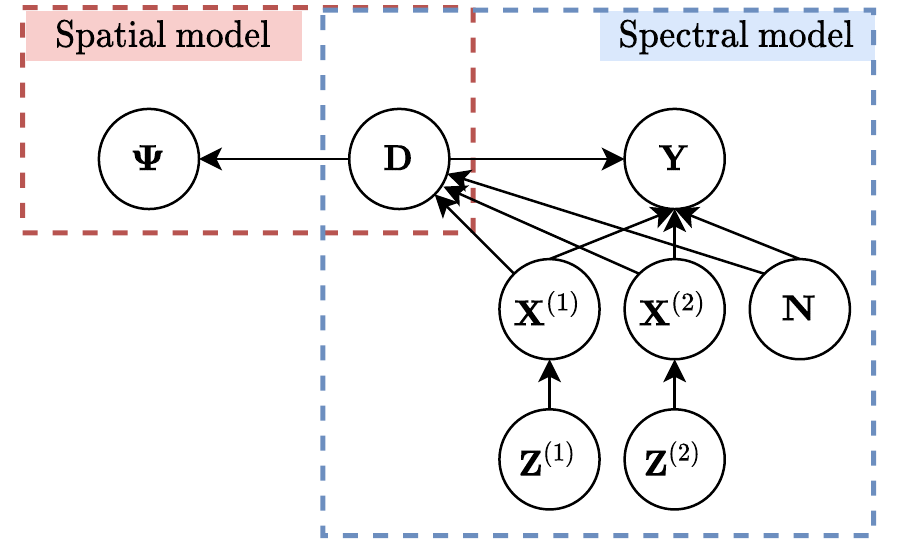}
    \caption{Graphical model of the integration of spatial and spectral model. The model applies to both DOLPHIN and the proposed model. The difference lies in the speech models $p(\mathbf{X}^{(1)}, \mathbf{Z}^{(1)})$ and $p(\mathbf{X}^{(2)}, \mathbf{Z}^{(2)})$.}
    \label{fig:graph_integr}
\end{figure}

\iffalse
\begin{table}
    \centering
    \caption{Notation.}
    \begin{tabular}{ll}
        \toprule
        $\boldsymbol{\Psi}$ & Spatial features \\
        $\mathbf{Y}$ & Mixed speech \\
        $\mathbf{X}_i$ & Speech from speaker $i$ \\
        $\mathbf{N}$ & Noise \\
        $\mathbf{D}$ & Index of dominant source \\
        \bottomrule
    \end{tabular}
\end{table}
\fi

In this section, we introduce the previously proposed approaches that we build upon, i.e. 1) spatial clustering with CACGMM, 2) DOLPHIN which combines spatial clustering with the factorial spectral model based on GMMs, and 3) integration of NN-based approach, PIT, with spatial clustering.

\subsection{Spatial clustering with Complex angular central Gaussian mixture model}
\label{subsec:spatial}
Spatial clustering approaches in general work on observed spatial features $\psi_{t,f}$ and aim to obtain soft assignments of the features at all time-frequency points to individual sources. For that, a mixture model is built, where one component corresponds to one source
\begin{align}
p(\boldsymbol{\psi}_{t,f}) &= \sum_{c} \pi_{c,t}\ p(\boldsymbol{\psi}_{t,f} |\ d_{t,f} = c),
\end{align}
where $c$ is the mixture component and $\pi_{c,t}$ is the mixture weight.
The assignment of the features to the components is represented by a latent categorical variable $\mathbf{D} = [d_{t,f}]_{t=1..T,f=1..F}$. Through the E-M algorithm, we can infer the model and estimate the soft assignments as posterior distribution $p(\mathbf{D}|\boldsymbol{\Psi})$, which indicates the probability of the source being active for each time-frequency bin and can thus be interpreted as a time-frequency mask.

There are different choices of spatial features and their models. Here, we use normalized observation vectors as spatial features and complex angular central Gaussian mixture model (CACGMM) \cite{ito:cacgmm,tyler1987statistical}. This combination has proven to be strong in recent work \cite{boeddeker2018front}. The normalized observation vectors are defined as
\begin{equation}
    \label{eq:spat}
    \boldsymbol{\psi}_{t,f} = \frac{\boldsymbol{\mathcal{Y}}_{t,f}}{|\boldsymbol{\mathcal{Y}}_{t,f}|}.
\end{equation}
Complex angular central Gaussian is defined as
\begin{align}
p(\boldsymbol{\psi}_{t,f} |\ d_{t,f} = c) &= \frac{(M-1)!}{2\pi^{M}\det \mathbf{B}_{c,f}} \frac{1}{( \boldsymbol{\psi}_{t,f}^{\mathsf H}  \mathbf{B}_{c,f}^{-1}  \boldsymbol{\psi}_{t,f})^{M}}, 
\end{align}
where $M$ is the number of channels (dimensionality of the spatial features) and $\mathbf{B}$ is the parameter of the distribution. For E-M algorithm in CACGMM, we refer the reader to previous work \cite{ito:cacgmm} and the implementation in \verb|pb_bss|\footnote{\url{https://github.com/fgnt/pb_bss}} \cite{drude:integration}.

\subsection{DOLPHIN}
\label{subsec:dolphin}
\begin{figure*}
    \centering
    \includegraphics[width=0.8\linewidth]{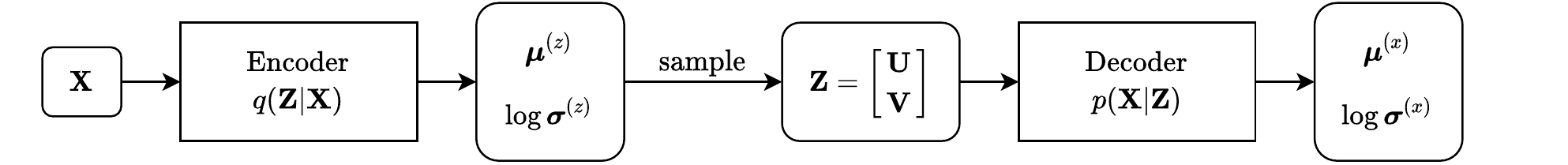}
    \caption{Scheme of the variational autoencoder used as a spectral model. Square blocks denote neural networks and round blocks their inputs and outputs.}
    \vspace{-1.5em}
    \label{fig:vae}
\end{figure*}
Spatial clustering exploits only spatial information in the signal and does not use the spectral structure of speech, such as the periodicity caused by the fundamental frequency of the voice, or temporal continuity of the speech signal. To improve upon this, integration of the spatial model with a spectral model was proposed \cite{nakatani2013dominance}, called DOLPHIN. There, the speech signal is modeled by a GMM
\begin{multline}
    p(\mathbf{x}_t) = \sum_{\mathbf{z}_t} p(\mathbf{z}_t) p(\mathbf{x}_t | \mathbf{z}_t) = \sum_c \pi^{(g)}_c \mathcal{N}(\mathbf{x}_t; \boldsymbol{\mu}^{(g)}_c, \mathrm{diag}(\boldsymbol{\sigma}_c^{(g)})),
\end{multline}
where $c$ is the mixture component and $\boldsymbol{\mu}^{(g)}_c, \mathrm{diag}(\boldsymbol{\sigma}_c^{(g)})$ are the mean and diagonal covariance matrix of the component $c$. The latent variable $\mathbf{Z} = [\mathbf{z}_t]_{t=1..T}$ models the assignment of time-frames to GMM mixture components.

The speech and noise models are pre-trained on a corpus of speech and noise and combined into the factorial model of the mixed signal, using the lifted max-model, which assumes that the log-magnitude STFT coefficients of the mixture is equal to that of the dominant source at each time-frequency bin\cite{rennie2010single}
\begin{align}
    d_{t,f} &= \argmax [n_{t,f}, x_{t,f}^{(1)}, x_{t,f}^{(2)}] \\
    y_{t,f} &= \begin{cases}
      n_{t,f} & \text{if } d_{t,f} = 0\vspace{0.5em}\\
      x_{t,f}^{(d_{t,f})} & \text{otherwise}\\
    \end{cases},  
\end{align}
where the latent variable $\mathbf{D} = [d_{t,f}]_{t=1..T,f=1..F}$ is the index of the dominant source at time-frequency bin $(t,f)$.
The lifted-max model has been previously used as an approximation of log-sum interaction \cite{rennie2010single}. In contrast with the max-model, the lifted max-model explicitly uses the index of the dominant source as a latent variable $\mathbf{D}$. This enables the integration with a spatial model by sharing $\mathbf{D}$, which is used to assign T-F point to mixture components in the spatial model. The integration of the factorial model with the spatial model is depicted in Figure \ref{fig:graph_integr}. 

The inference in DOLPHIN iteratively estimates the latent variables $\mathbf{D}$ (the time-frequency masks of the sources), latent variables $\mathbf{Z}$ (spectral patterns of the speech sources) and parameters of the spatial model ($\mathbf{B}$ in our case). The procedure is analogous to inference in our proposed model, described in detail in section \ref{subsec:inference}, with the difference of the estimation of $q(\mathbf{Z})$, which has an analytical solution for DOLPHIN with GMM. For details, we refer to the original paper \cite{nakatani2013dominance}.

\subsection{Combination of PIT and spatial clustering}
\label{subsec:pit_spatial}
Neural networks became widely used for speech separation with the emergence of techniques such as deep clustering \cite{hershey2016deep} and PIT \cite{yu2017permutation}. In these, neural networks are discriminatively trained models that map the mixed speech to the individual sources. These models are single-channel models, but they can be integrated with multi-channel processing in several ways. The naive way consists of simply using the time-frequency mask provided by the separation approach to design a beamformer \cite{higuchi2017deep}. A tighter integration was proposed in \cite{nakatani2017integrating}, where T-F masks produced by PIT were used as priors $\boldsymbol{\pi}$ for the source affiliations $\mathbf{D}$ in the spatial clustering model (Section \ref{subsec:spatial}). This way, the spatial model is pulled towards the solution provided by PIT, which uses spectral information.

Inference in the integrated model is then analogous to inference in the spatial model by itself, with the difference of the mixture weights being fixed to the T-F masks estimated by PIT.

\section{Proposed method}
\label{sec:proposed}
\vspace{-0.5em}
In this work, we follow the integration of the spectral and spatial model used in DOLPHIN, however, we replace the GMM speech model by a variational autoencoder (VAE), which is a neural network-based generative model. We believe that VAE has better modeling power than GMM due to its non-linear transformation and possibility to model correlations over time. In this section, we explain the construction and training of the VAE as a spectral model, revise the structure of the full integrated model, and introduce how the inference in the integrated model works.

\vspace{-0.4em}
\subsection{Variational autoencoder as spectral model}
\vspace{-0.1em}
Among the different models rising in the field in deep generative modeling, variational auto-encoder \cite{kingma2013auto,rezende2014stochastic} has become probably the most popular in speech processing \cite{hsu2017unsupervised,gburrek2019unsupervised}. Variational auto-encoder is a deep latent variable model, where the probability density of the observation depends on the output of a highly non-linear transformation of the latent variable, where the transformation is modeled by a neural network (known as a decoder). The parameters of the model are trained by maximizing the variational lower bound of the log-likelihood of the observed data. To make the training fast, another neural network (known as encoder) is used for inference of the approximate posterior of the latent variables given observed data \cite{rezende2014stochastic}.

In our work, we use VAE to model single-speaker speech in the log-magnitude STFT domain. We use a (slightly modified) model from \cite{gburrek2019unsupervised}, which was inspired by \cite{hsu2017unsupervised}. In this model, part of the latent variables is trained to model speaker variability. As the speaker variability is quite an important factor for speech separation, we consider explicitly disentangling this information in the latent space useful. We denote $\mathbf{U} = [\mathbf{u}_t]_{t=1..T}$ the speaker-independent part of the latent variable, $\mathbf{V} = [\mathbf{v}_t]_{t=1..T}$ the speaker-dependent part, and $\mathbf{Z} = [\mathbf{z}_t]_{t=1..T}$ their concatenation\footnote{To emphasize the common structure of DOLPHIN with GMM and DOLPHIN with VAE, we slightly abuse notation and use $\mathbf{Z}$ as the latent variable in both. In Section \ref{sec:proposed}, $\mathbf{Z}$ refers to VAE latent variable as defined in Eq. (\ref{eq:vae_z}).}. The model of the latent variable is as follows:
\begin{align}
p(\mathbf{u}_t) &= \mathcal{N}(\mathbf{u}_t; \mathbf{0}, \mathbf{I}) \\
p(\mathbf{v}_t) &= \mathcal{N}(\mathbf{v}_t; \boldsymbol{\mu}_s^{(spk)}, \mathbf{I}) \\
\label{eq:vae_z} \mathbf{z}_t &= [\mathbf{u}_t, \mathbf{v}_t], \\
p(\mathbf{z}_t) &= \mathcal{N}(\mathbf{z}_t; [\mathbf{0}, \boldsymbol{\mu}_s^{(spk)}], \mathbf{I}), 
\end{align}
where $\boldsymbol{\mu}_s^{(spk)}$ is a speaker-dependent mean of speaker $s$. The speech signal is then modeled as a normal distribution with parameters dependent on neural network transformation of the latent variable
\begin{equation}
    p(\mathbf{x}_t |\ \mathbf{Z}) = \mathcal{N}(\mathbf{x}_t; \boldsymbol{\mu}_t^{(x)}(\mathbf{Z}), \mathrm{diag}(\boldsymbol{\sigma}_t^{(x)}(\mathbf{Z}))),
\end{equation}
where $\boldsymbol{\mu}_t^{(x)}(\cdot)$ and $\boldsymbol{\sigma}_t^{(x)}(\cdot)$ are functions modeled by the VAE decoder.  
The variational posterior $q(\mathbf{Z})$ infered by the encoder is assumed to be Gaussian with parameters $\boldsymbol{\mu}^{(z)}_t$, $\boldsymbol{\sigma}^{(z)}_t$ for every time frame $t$
\begin{equation}
    \label{eq:qz_vae}
    q(\mathbf{Z}) = \prod_t q(\mathbf{z}_t) = \prod_t \mathcal{N}(\mathbf{z}_t; \boldsymbol{\mu}^{(z)}_t, \mathrm{diag}(\boldsymbol{\sigma}^{(z)}_t)).
\end{equation}
The scheme of the described VAE is depicted in Figure \ref{fig:vae}. The parameters of the encoder, decoder, and speaker means are trained with an objective function, which combines variational lower-bound with speaker classification objective. The speaker classification objective encourages $\mathbf{V}$ to model speaker information. The full objective is:
\vspace{-0.8em}
\begin{multline}
\mathcal{J}_{\text{VAE}} = \overbrace{\mathbb{E}_{\mathbf{Z} \sim q(\mathbf{Z})}[\ln p(\mathbf{X}|\mathbf{Z})] - \text{KL}(q(\mathbf{Z})||p(\mathbf{Z}))}^\text{\strut variational lower-bound} +\\+ \underbrace{\sum_s \ln p(s | \boldsymbol{\mu}_s^{(spk)})}_\text{\strut speaker classification}.
\end{multline}
Note that the speaker classification objective differs from \cite{gburrek2019unsupervised,hsu2017unsupervised}, where the aim was to learn to disentangle the speaker information in an unsupervised way. Here, we assume having the speaker identities during training, and  we can thus directly use a supervised objective.

\vspace{-0.25em}
\subsection{Full integrated model}
\vspace{-0.5em}
To compose the full integrated model, we follow the model of DOLPHIN. The factorial model of mixed signal is composed of the VAE speech model for each speaker, and a noise model. In our work, we used a simple Gaussian model for noise
\begin{equation}
    p(\mathbf{n}_t) = \mathcal{N}(\mathbf{n}_t; \boldsymbol{\mu}^{(n)}, \mathrm{diag}(\boldsymbol{\sigma}^{(n)})).
\end{equation}
As there are two VAE speech models in the integrated model, for two speakers, we distinguish between the variables and distribution parameters by upper index $^{(i)}$, e.g. $\mathbf{Z}^{(1)} = [\mathbf{z}^{(1)}_t]_{t=1..T}$ for the latent variable of the first speaker in the mixture, and $\boldsymbol{\mu}^{(x,1)}_t$ for the mean of $p(\mathbf{x}_t^{(1)} | \mathbf{Z}^{(1)})$.

The interaction model is the lifted-max model introduced in Section \ref{subsec:dolphin}. As for conventional DOLPHIN, the integration with the spatial model is done through the variable $\mathbf{D}$ used as mixture component affiliation in the spatial model, and index of dominant source in the spectral model. The graphical model of the full integrated model is depicted in Figure \ref{fig:graph_integr}. 
\iffalse
The complete data likelihood has the following form
\begin{multline}
    p(\boldsymbol{\Psi}, \mathbf{Y}, \mathbf{D}, \mathbf{X}^{(1)}, \mathbf{X}^{(2)}, \mathbf{Z}^{(1)}, \mathbf{Z}^{(2)}, \mathbf{N}) =\\ =  p(\boldsymbol{\Psi} | \mathbf{D}) p (\mathbf{Y}, \mathbf{D} | \mathbf{Z}^{(1)}, \mathbf{Z}^{(2)}, \mathbf{N}) p(\mathbf{Z}^{(1)})p(\mathbf{Z}^{(2)})p(\mathbf{N})
\end{multline}
\fi

\vspace{-0.25em}
\subsection{Inference}
\label{subsec:inference}
\vspace{-0.5em}

The speech and the noise models are pre-trained in advance on non-overlapped speech and noise data. During the inference time, given an observed mixture, we want to obtain an estimate of $p(\mathbf{D}|\mathbf{Y},\boldsymbol{\Psi})$, i.e. time-frequency masks of the individual sources. For that, we employ Variational Bayes inference, assuming the approximate posterior to factorize as follows
\begin{equation} \label{eq1}
\begin{split}
q(\mathbf{D}, \mathbf{Z}^{(1)}, \mathbf{Z}^{(2)}) &= q(\mathbf{D})q(\mathbf{Z}^{(1)})q(\mathbf{Z}^{(2)})  \\ &= \prod_{t,f} q(d_{t,f})\  q(\mathbf{Z}^{(1)})q(\mathbf{Z}^{(2)}).
\end{split}
\end{equation}
Further, the approximate posterior over the latent variables of the spectral models, $q(\mathbf{Z}^{(i)})$, are assumed to be Gaussian as in training of the VAE in~Eq.~(\ref{eq:qz_vae})
\begin{align}
q(\mathbf{Z}^{(1)}) = \prod_t q(\mathbf{z}^{(1)}_t) &= \prod_t  \mathcal{N}(\mathbf{z}^{(1)}_t; \boldsymbol{\mu}^{(z,1)}_t, \mathrm{diag}(\boldsymbol{\sigma}^{(z,1)}_t)) \\
q(\mathbf{Z}^{(2)}) = \prod_t q(\mathbf{z}^{(2)}_t) &= \prod_t \mathcal{N}(\mathbf{z}^{(2)}_t; \boldsymbol{\mu}^{(z,2)}_t, \mathrm{diag}(\boldsymbol{\sigma}^{(z,2)}_t)). 
\end{align}
The parameters of $q(\mathbf{Z}^{(1)})$, $q(\mathbf{Z}^{(2)})$ are thus $\theta^{(z,1)} = (\boldsymbol{\mu}^{(z,1)}, \boldsymbol{\sigma}^{(z,1)})$ and $\theta^{(z,2)} = (\boldsymbol{\mu}^{(z,2)}, \boldsymbol{\sigma}^{(z,2)})$.

During the inference, we need to infer 1) the approximate posterior $q(\mathbf{D})$, 2) the parameters $\theta^{(z,1)},\theta^{(z,2)}$ of the approximate posterior $q(\mathbf{Z}^{(1)})q(\mathbf{Z}^{(2)})$ and 3) the parameters of the spatial model $\mathbf{B}$. These three inference steps are performed iteratively. In the beginning of the inference, the parameters may be initialized randomly. For faster convergence, we initialize $q(\mathbf{Z}^{(1)})q(\mathbf{Z}^{(2)})$ by forwarding the observed mixed speech through the VAE encoder. 

\subsubsection{Inference of $q(\mathbf{D})$}
\vspace{-0.3em}
Approximate posterior distribution $q(\mathbf{D})$ represents the T-F masks of speech and noise sources. Estimation of the masks for source $k$ should consider how well the observations at different T-F points fit under the spatial and spectral model of the source $k$. For the spatial model, it is the $k$-th component of the mixture. For the spectral model, it is either the noise model or VAE speech model with the current estimate of its latent variables $q(\mathbf{Z}^{(i)})$. 

Formally, the update follows Variational Bayes, we thus maximize the variational lower-bound. This has an analytical solution (following Eq. 10.9 in \cite{bishop2006pattern})
\vspace{-0.8em}
\begin{multline}
\label{eq:qd_up}
\ln q(d_{t,f} = k) = \overbrace{\ln p(\boldsymbol{\psi}_{t,f} | d_{t,f} = k)}^\text{spatial model} + \\ \underbrace{\mathbb{E}_{\mathbf{Z}^{(1)}, \mathbf{Z}^{(2)} \sim q(\mathbf{Z}^{(1)}, \mathbf{Z}^{(2)})}[\ln p(y_{t,f}, d_{t,f} = k | \mathbf{Z}^{(1)}, \mathbf{Z}^{(2)})]}_\text{spectral model} + C
\end{multline}
where $C$ is constant w.r.t $d_{t,f}$. We obtain the constant $C$ by ensuring that $\sum_k q(d_{t,f} = k) = 1$. The expectation is approximated by sampling $N_z$ samples from $q(\mathbf{Z}^{(1)}, \mathbf{Z}^{(2)})$. The value inside the expectation expresses how likely the source $k$ is dominant and has the value of the observed mixed speech $y_{t,f}$. For example of $k=1$, this can be expressed as 
\vspace{-0.5em}
\begin{multline}
\label{eq:qd_detail}
    \ln p(y_{t,f}, d_{t,f} = 1 | \mathbf{Z}^{(1)}, \mathbf{Z}^{(2)}) = \\\ln p(x^{(1)}_{t,f} = y_{t,f} | \mathbf{Z}^{(1)}) + \ln p(x^{(2)}_{t,f} < y_{t,f} | \mathbf{Z}^{(2)}) \\+ \ln p(n_{t,f} < y_{t,f}) = \\ = \ln \mathcal{N}(y_{t,f}; \mu_{t,f}^{(x,1)}, \sigma_{t,f}^{(x,1)}) + \ln \Phi(y_{t,f}; \mu_{t,f}^{(x,2)}, \sigma_{t,f}^{(x,2)}) \\+ \ln \Phi(y_{t,f}; \mu_{t,f}^{(n)}, \sigma_{t,f}^{(n)}),
\end{multline}
where $\Phi(\cdot)$ is the cumulative normal distribution. This solution can be also found in previous literature on lifted-max model \cite{rennie2010single, nakatani2013dominance}.

\subsubsection{Inference of $q(\mathbf{Z}^{(1)})q(\mathbf{Z}^{(2)})$}
\vspace{-0.3em}
Given the current estimate of the masks $q(\mathbf{D})$, we estimate the spectral characteristics of the speech of the two speakers, represented as latent variables of the VAE speech models $\mathbf{Z}^{(1)},\mathbf{Z}^{(2)}$. These should be optimized in such a way that they correspond to the current estimate of $q(\mathbf{D})$ (source estimated as likely dominant should be likely dominant) and also represent proper speech signal (they are close to the prior distribution of $\mathbf{Z}^{(1)},\mathbf{Z}^{(2)}$).

Formally, we optimize the variational lower bound. In this case, the optimal value for the parameters of $q(\mathbf{Z}^{(1)}, \mathbf{Z}^{(2)})$ cannot be found analytically, we thus employ a few steps of gradient ascent optimization of the lower-bound  (following Eq. 10.6 in \cite{bishop2006pattern})
\begin{multline}
\label{eq:qz_lb}
\mathbb{E}_{\mathbf{Z}^{(1)}, \mathbf{Z}^{(2)} \sim q(\mathbf{Z}^{(1)}, \mathbf{Z}^{(2)})} \mathbb{E}_{\mathbf{D} \sim q(\mathbf{D})} [\ln p(\mathbf{Y}, \mathbf{D} | \mathbf{Z}^{(1)}, \mathbf{Z}^{(2)})] - \\ \lambda\ KL(q(\mathbf{Z}^{(1)}, \mathbf{Z}^{(2)})|| p(\mathbf{Z}^{(1)}, \mathbf{Z}^{(2)})),
\end{multline}
where $\lambda$ is a hyper-parameter weighting two parts of the objective. The theoretically correct value of $\lambda$ is 1, it may however help to improve the inference to choose a different value. The expectation over $\mathbf{Z}$ is approximated by Monte-Carlo samples. The expectation over $\mathbf{D}$ can be further factorized into individual T-F points. The KL divergence can be computed analytically as both $q(\mathbf{Z}^{(1)}, \mathbf{Z}^{(2)})$ and $p(\mathbf{Z}^{(1)}, \mathbf{Z}^{(2)})$ are Gaussian distributed. The speaker means $\boldsymbol{\mu}_s^{(spk)}$, which are part of $p(\mathbf{Z}^{(i)})$ are set to their optimal value w.r.t. the lower-bound, which is an average of $\boldsymbol{\mu}^{(z,i)}_t$ over time. To compute gradient w.r.t. the parameters of $q(\mathbf{Z}^{(1)}, \mathbf{Z}^{(2)})$ that we sample from, we employ the reparametrization trick~\cite{kingma2013auto}.

\subsubsection{Inference of spatial model parameters}
\vspace{-0.3em}
Finally, the parameters of the spatial model can be updated using an M-step of the E-M algorithm following the inference procedure in the spatial model itself and using the current estimate of $q(\mathbf{D})$. For details on M-step on CACGMM, we refer to \cite{ito:cacgmm}.

\subsubsection{Inference summary}
\vspace{-0.3em}
For clarity, we summarize the inference process in the following algorithm. Details on the initialization steps are given in experimental settings in Section \ref{subsec:settings}.
\vspace{-0.3em}
\begin{algorithm}
\caption{Inference in the proposed integrated model.}
\begin{enumerate}
    \item Initialize $q(\mathbf{D})$.
    \item Initialize parameters $\theta^{z,1}$, $\theta^{z,2}$ of $q(\mathbf{Z}^{(1)})$, $q(\mathbf{Z}^{(2)})$.
    \item Iterate the following for $N_{\text{iter}}$ iterations.
    \begin{enumerate}
        \item Update $q(\mathbf{D})$ using Eq. (\ref{eq:qd_up}), (\ref{eq:qd_detail}).
        \item Update parameters $\theta^{z,1}$, $\theta^{z,2}$ of $q(\mathbf{Z}^{(1)})$, $q(\mathbf{Z}^{(2)})$ \\by iterating the following for $N_{\text{up}}$ iterations.
        \begin{enumerate}
            \item Sample $\mathbf{Z}^{(1)}$, $\mathbf{Z}^{(2)}$ from $q(\mathbf{Z}^{(1)})$, $q(\mathbf{Z}^{(2)})$.
            \item Forward $\mathbf{Z}^{(1)}$, $\mathbf{Z}^{(2)}$ through VAE decoder.
            \item Compute gradients of Eq. (\ref{eq:qz_lb}) w.r.t. $\theta^{z,1}$, $\theta^{z,2}$ by back-propagating through the VAE decoder.
            \item Update $\theta^{z,1}$, $\theta^{z,2}$ with gradient ascent.
        \end{enumerate}
    \item Update the parameters $\mathbf{B}$ of the spatial model \cite{ito:cacgmm}.
    \end{enumerate}
\end{enumerate}
\end{algorithm}

\begin{figure*}[t]
    \centering
    \includegraphics[width=0.9\linewidth]{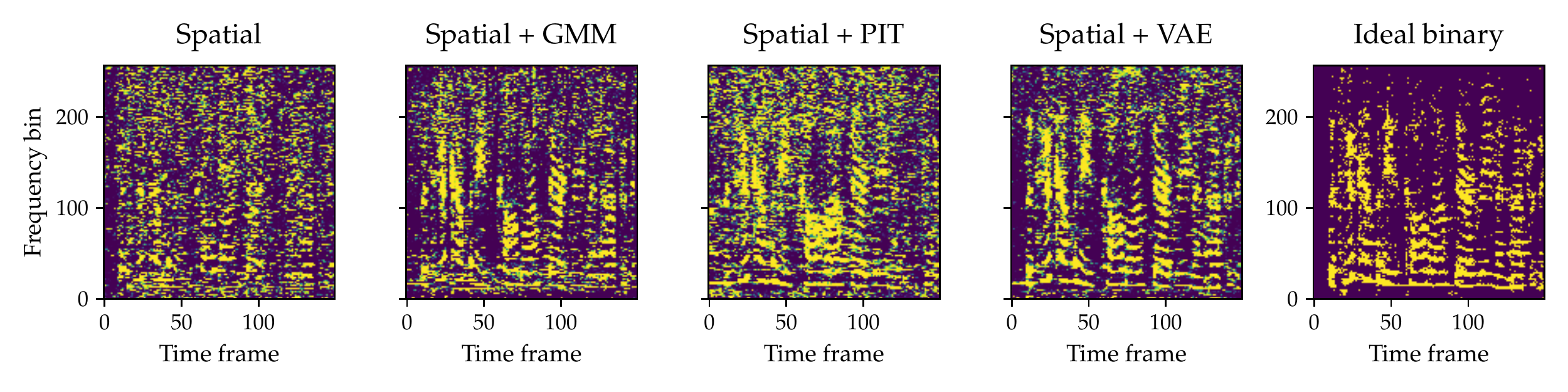}
    \vspace{-0.5em}
    \caption{Time-frequency masks obtained by the tested methods, and oracle ideal binary mask. The color scale in all masks ranges from 0 to 1.}
    \label{fig:masks}
    \vspace{-0.5em}
\end{figure*}

\vspace{-0.7em}
\section{Experiments}
\vspace{-0.5em}
\subsection{Datasets}
\vspace{-0.3em}
As a dataset for our experiments, we used the spatialized version of WSJ0-2mix\footnote{\url{http://www.merl.com/demos/deep-clustering}} sampled at 8kHz, which was introduced in \cite{wang2018multi}. The original WSJ0-2mix dataset \cite{hershey2016deep} was created by mixing utterances of WSJ0 \cite{paul1992design} with SNR of one source to the other of -5-5 dB. It consists of training, validation, and evaluation sets of 20000, 5000, and 3000 utterances, respectively. The evaluation set contains a disjoint set of speakers from training and validation. The spatialized version was created by using artificially generated room impulse responses with 8 microphones, randomized setting of microphone array size (15-25 cm), T60s (0.2 - 0.6 s), and speaker distance to the array (1.3m with 0.4m standard deviation). In our work, we additionally added a white diffuse noise to the data, to analyze the possibility of the methods to adapt to different noise conditions. We created 5 different conditions with SNRs of the mixed speech to the noise of 15-20, 10-15, 5-10, 0-5, and -5-0 dB.

\subsection{Settings}
\label{subsec:settings}
\begin{description}
\itemsep-0.2em 
\item{\textbf{Signal representation}} All experiments use Short-time Fourier transform (STFT) with a window of 512 samples (64 ms), a shift of 128 samples (16ms)  and a Hanning window. For VAE and GMM, we use the log-magnitude of STFT coefficients, for PIT training we use the magnitude of the STFT coefficients.
\item{\textbf{GMM settings}} We use 256-component GMM. The GMM was trained by progressively splitting the mixture by doubling the number of components. We used 1-4-4-10-10-10-10-10-10 E-M iterations for each stage with 1-2-4-8-16-32-64-128-256 components.
\item{\textbf{PIT architecture and training}} For PIT, we used an available implementation\footnote{\url{https://github.com/funcwj/uPIT-for-speech-separation}}. The architecture consists of 3 BLSTM layers with hidden size 896 and ReLU activation functions. The training uses Adam optimizer with learning rate 1e-3, momentum 0.9, weight decay 1e-5, dropout rate 0.5. We train the PIT network on mixtures with 15-20 dB noise and use three output layers (for two speakers and noise). The outputs are T-F masks and the objective function is the phase-sensitive mean-squared error of the masked mixture \cite{erdogan2015phase}.
\item{\textbf{VAE architecture and training}} The architecture of the encoder and decoder roughly follows \cite{gburrek2019unsupervised}. The encoder has 5 1-d convolutional layers and 4 feed-forward layers with hidden dimensionality of 512. The kernel size in all convolutional layers is 3, the stride is 1-1-2-1-1. ReLU activations are used at all layers. Every layer which does not change temporal resolution has an additional residual connection. The final feed-forward layer outputs the parameters of $q(\mathbf{Z})$: $\boldsymbol{\mu}^{(u)}, \boldsymbol{\mu}^{(v)}$ and $\log\boldsymbol{\sigma}^{(u)}, \log\boldsymbol{\sigma}^{(v)}$, which all have a dimension of 20. The dimension of the input to the encoder is 257. The decoder has the same architecture as the encoder, but all convolutions are replaced with transposed convolutions. The dimension of the input to the decoder is 40 (dimension of $\mathbf{u}$ + dimension of $\mathbf{v}$). The output of the decoder are the parameters of $p(\mathbf{X}|\mathbf{Z})$: $\boldsymbol{\mu}^{(x)}, \log\boldsymbol{\sigma}^{(x)}$, which both have a dimension of 257. The VAE was trained for 100 epochs with Adam optimizer, learning rate 1e-4, and gradient norm clipping 10.
\item{\textbf{Inference}} We initialize both $q(\mathbf{Z}^{(1)})$ and $q(\mathbf{Z}^{(2)})$ by forwarding the mixed speech through encoder of the VAE, and $q(\mathbf{D})$ randomly from uniform distribution with subsequent normalization. The inference runs for $N_{\text{iter}} = 100$ iterations (this is common for all tested methods). For updating parameters of $q(\mathbf{Z}^{(1)})q(\mathbf{Z}^{(2)})$, we used $N_{\text{up}} = 5$ updates in each iteration with learning rate 1e-3. We set the weight of the KL divergence part of the objective in Eq.~(\ref{eq:qz_lb}) to 10, which was optimal for validation set. The expectations over $q(\mathbf{Z}^{(1)})q(\mathbf{Z}^{(2)})$ are approximated using one sample.
\item{\textbf{Beamforming}} To obtain the separated sources, we use a mask-based MVDR beamformer \cite{heymann2016neural} as defined in \cite{Souden2010MVDR,Erdogan2016MVDR}.
\end{description}

\subsection{Results}
The results of the experiments are summarized in Table \ref{tab:results}. We report signal-to-distortion-ratio\footnote{\url{https://github.com/craffel/mir_eval}}\cite{raffel2014mir_eval} (SDR) improvements over the original mixtures. The SDR values of the mixtures for each condition are shown in the header of the table. All experiments use mask-based beamforming to obtain the separated signals. The tested methods are \textit{spatial} clustering (Section \ref{subsec:spatial}), \textit{spatial+GMM} (Section \ref{subsec:dolphin}), \textit{spatial+PIT} (Section \ref{subsec:pit_spatial}) and proposed \textit{spatial+VAE} (Section \ref{sec:proposed}). In addition, we also show results of using the masks provided by PIT directly to estimate beamformer without using spatial clustering, denoted as \textit{PIT-bf}, which has been widely used \cite{yoshioka2018recognizing,higuchi2017deep}. Note that PIT network is trained on noise condition with SNR 15-20 dB, in order to analyze the effect of mismatched noise conditions. Lastly, \textit{Ideal binary mask} shows an oracle result.

First, let us focus on the comparison between between \textit{spatial}, \textit{spatial+GMM} and \textit{spatial+VAE} (rows 1,2,4). The combination with GMM is consistently better than spatial clustering by itself which confirms the advantage of using the spectral model, as proposed in DOLPHIN \cite{nakatani2013dominance}. Replacing GMM with VAE spectral model (as proposed here) significantly improves the performance, showing that VAE is a stronger model of the speech signal. 

Second, we compare \textit{spatial+PIT} with \textit{spatial+VAE} on matched condition (rows 3,4, column 1). The results show comparable performance of both methods. This shows the validity of our approach and the potential of using deep generative modeling for speech separation, in contrast with discriminatively trained networks, which are today primarily used. Interestingly, no part of our model was trained on mixed speech, but nevertheless, it can achieve the performance of PIT which was trained directly for separation. Note that the commonly used beamforming based on PIT masks 
(row 5) under-performs the integrated \textit{spatial+PIT} model by a significant margin.

Finally, we explore the possibility of adapting the \textit{spatial+VAE} model to a different noise condition. For every condition, we estimated the noise model parameters ($\boldsymbol{\mu}^{(n)}$, $\boldsymbol{\sigma}^{(n)}$) on the noise signals of corresponding SNRs. The rest of the model remains fixed. We can see the effect by comparing with \textit{spatial+PIT} (rows 3,4), which is not adapted. As the mismatch grows, the margin between \textit{PIT} and \textit{VAE} increases. This shows that the model is successfully adapting by re-training the noise model. In practice, we could for example estimate the noise model on non-speech parts of recordings and possibly vary the complexity of the noise model based on how much data we have available.

Figure \ref{fig:masks} shows the time-frequency masks obtained with the spatial model and combination of spatial and spectral models with GMM, PIT, and VAE. We also show the ideal binary mask. Looking at the figure, we observe that the combination of spatial and spectral models lead to masks that better represent the harmonic structure of speech. Comparing the masks obtained with GMM and with VAE, we see the better modeling capability of the VAE, especially at lower frequencies. Moreover, compared to spatial+PIT masks, the VAE mask is less noisy and much closer to the ideal binary mask. This improved mask estimation performance translates in better speech enhancement performance after beamforming.

\vspace{-0.4em}
\section{Related works}
\begin{table}[t]
\centering
\caption{Signal-to-distortion ratio gains (dB) of the tested methods on the spatialized WSJ0-2mix corpus with different levels of added noise.}
\vspace{-0.6em}
\begin{tabular}{@{}llccccc@{}}\cmidrule[\heavyrulewidth]{1-7}
 &noise SNR [dB] & 15-20 & 10-15 & 5-10 & 0-5 & -5-0 \\
&mixture SDR [dB] & -0.5 & -0.9 & -1.8 & -3.8 & -6.9 \vspace{0.5em}\\
\cmidrule{1-7}
\multicolumn{7}{c}{$\Delta$ SDR [dB]} \\
\cmidrule{1-7}
%&$\Delta$ SDR [dB] &&&&& \\\cmidrule{2-7}
{\small 1}&Spatial & 11.6 & 10.0 & 8.1 & 6.3 & 5.5 \\
{\small 2}&+ GMM (DOLPHIN) & 12.1 & 10.6 & 8.8 & 7.3 & 6.8 \\
{\small 3}&+ PIT & 12.6 & 11.2 & 9.6 & 7.7 & 7.0 \\
{\small 4}&+ VAE (Proposed) & \textbf{12.7} & \textbf{11.4} & \textbf{10.0} & \textbf{8.9} & \textbf{8.1} \\
{\small 5}&PIT-bf & 10.8 & 9.7 & 8.2& 6.2&3.3 \\ \cmidrule{1-7}
{\small 6}&Ideal binary mask & 13.5 & 12.2 & 10.9 & 10.1 & 10.1 \\
\cmidrule[\heavyrulewidth]{1-7}
\end{tabular}
\label{tab:results}
\vspace{-1em}
\end{table}
\vspace{-0.6em}
The concept of using a variational auto-encoder as the speech model for speech enhancement or separation has been explored in several previous works. For single-channel speech enhancement, authors in \cite{bando2018statistical,leglaive2018variance,pariente2019statistically} use VAE speech model and low-rank model of noise together with Gibbs sampling to estimate clean speech spectra from noisy spectrogram. This was further extended to multi-channel scenario \cite{sekiguchi2018bayesian,leglaive2019semi,nugraha2020flow} by integrating the models into multi-channel nonnegative matrix factorization (MNMF) framework. 

Similar methods have been explored for multi-channel speech separation in \cite{seki2019generalized,li2019fast}. The main difference to our proposal lies in the multi-channel modeling --- authors in \cite{seki2019generalized,li2019fast} extend the MNMF framework, while we extend DOLPHIN, which is based on spatial clustering. The works are thus complementary to each other and pursue similar ideas in different settings, which both have their pros and cons. The MNMF+VAE approach uses linear spectra which can be advantageous because the mixture can be accurately modeled by the sum of component signals. Further, due to their use of Multi-channel Wiener filter (MWF), the optimization and estimation use the same objective. On the other hand, the use of CACGMM and spatial clustering has been shown very effective for far-field ASR and is computationally much less demanding than MWF. The MNMF+VAE based methods have been so far tested on a relatively small dataset with a limited number of speakers, direct comparison is thus difficult. 

Several ideas from the previous works could be also incorporated into our proposal in future extensions, such as the low-rank model of noise, re-using the VAE encoder for faster inference \cite{pariente2019statistically,li2019fast} or more flexible speech modeling with flows \cite{nugraha2020flow}.

\vspace{-0.1em}
\section{Conclusion}
\vspace{-0.1em}
In this paper, we combined a VAE spectral model of the speech signal with CACGMM spatial clustering for multi-channel speech separation. This extends previously proposed DOLPHIN method which uses GMM as the spectral model. We showed that VAE, as a more flexible model, outperforms GMM on all tested conditions. Furthermore, we showed comparable performance to PIT integrated with spatial clustering and the possibility of adaptation to different noise conditions enabled by an explicit noise model. Future work includes investigations on datasets with more complex noisy conditions and adaptation of the noise models within the inference step. The method could be also extended on the model side, e.g. with more structured latent space or a different choice of deep generative models. The code for the method is released on Github\cite{zmolikovagithub:vaedolphin}.

% References should be produced using the bibtex program from suitable
% BiBTeX files (here: strings, refs, manuals). The IEEEbib.bst bibliography
% style file from IEEE produces unsorted bibliography list.
% -------------------------------------------------------------------------
\bibliographystyle{IEEEbib}
\bibliography{strings,refs}

\end{document}